\documentclass[a4paper,11pt]{article}

\begin{document}
\topmargin 0pt \oddsidemargin 0mm

\renewcommand{\thefootnote}{\fnsymbol{footnote}}

\begin{titlepage}
\begin{flushright}
gr-qc/0612089
\end{flushright}

\vspace{5mm}
\begin{center}
{\Large \bf Thermodynamic Behavior of Field Equations for $f(R)$
Gravity} \vspace{12mm}

{\large M. Akbar\footnote{Email address: akbar@itp.ac.cn} and
Rong-Gen Cai\footnote{Email address: cairg@itp.ac.cn }}\\
\vspace{6mm} { \em Institute of Theoretical Physics, Chinese
Academy of Sciences,\\
 P.O. Box 2735, Beijing 100080, China}
\vspace{10mm}

\end{center}
\vspace{20mm} \centerline{{\bf{Abstract}}}
 \vspace{5mm}
Recently it has shown that Einstein's field equations can be
rewritten into a form of the first law of thermodynamics both at
event horizon of static spherically symmetric black holes and
apparent horizon of Friedmann-Robertson-Walker (FRW) universe, which
indicates intrinsic thermodynamic properties of spacetime horizon.
In the present paper we deal with the so-called $f(R)$ gravity,
whose action is a function of the curvature scalar $R$. In the setup
of static spherically symmetric black hole spacetime, we find that
at the event horizon, the field equations of $f(R)$ gravity can be
written into a form $dE = TdS - PdV + Td\bar{S}$, where $T$ is the
Hawking temperature and $S=Af'(R)/4G$ is the horizon entropy of the
black hole, $E$ is the horizon energy  of the black hole, $P$ is the
radial pressure of matter, $V$ is the volume of black hole horizon,
and $d\bar S$ can be interpreted as the entropy production term due
to nonequilibrium thermodynamics of spacetime. In the setup of FRW
universe, the field equations  can also be cast to a similar form,
$dE=TdS +WdV +Td\bar S$, at the apparent horizon, where
$W=(\rho-P)/2$, $E$ is the energy of perfect fluid with energy
density $\rho$ and pressure $P$ inside the apparent horizon.
Compared to the case of Einstein's general relativity, an additional
term $d\bar S$ also appears here. The appearance of the additional
term is consistent with the argument recently given by Eling {\it et
al.} (gr-qc/0602001) that the horizon thermodynamics is
non-equilibrium one for the $f(R)$ gravity.

\end{titlepage}

\newpage
\renewcommand{\thefootnote}{\arabic{footnote}}
\setcounter{footnote}{0} \setcounter{page}{2}

\section{Introduction}\label{sec1}
The four laws of black hole mechanics within Einstein's theory of
general relativity are very closely analogous to the four laws of
the usual thermodynamics~\cite{a3}. The quantities that provide a
basis for the black hole thermodynamics are the Hawking temperature
$T = \kappa / 2\pi$, entropy $S = A / 4G$ and energy $E$, where
$\kappa$ is the surface gravity, $A$ is the area of the event
horizon and $E = M$ is the mass of the black hole. The Hawking
temperature together with black hole entropy is connected through
the identity $TdS = dE$ and is usually called the first law of black
hole thermodynamics \cite{a3,a1,a2}. In general, the first law of
black hole thermodynamics is related with the energy change, $dE$
when a black hole varies from one stationary state to another by
\begin{equation}\label{1}
dE = \frac{\kappa}{8\pi}dA + {\rm work\ terms,} ~~~or~~~dE = TdS +
{\rm work\ terms}.
\end{equation}
The ``work terms" are given differently depending upon the type of
the black holes. For Kerr-Newman black hole family, the first law of
black hole thermodynamics states that the differences in mass $M$,
the horizon area $A$, the angular momentum $J$, and the electric
charge $Q$ of two nearby black holes are connected by $dM = TdS +
\Omega dJ + \Phi dQ$, where $\Omega = \frac{\partial M}{\partial J}$
is the angular velocity and $\Phi = \frac{\partial M}{\partial Q}$
is the electric potential. The first law of black hole
thermodynamics seemingly indicates that there should be some
relationship between thermodynamics and Einstein field equations
because black hole solutions are derived from the Einstein equations
and the geometric quantities (horizon area, surface gravity) of the
spacetime metric are related with thermodynamical quantities
(entropy, temperature) of the thermal system. In this regard,
Jacobson \cite{a8} is the first one to explore such an relation. He
found that it is indeed possible to derive the Einstein's equations
from the proportionality of entropy to the horizon area together
with the fundamental relation $\delta Q= TdS$, assuming the relation
holds for all local Rindler causal horizons through each spacetime
point. Here $\delta Q$ and $T$ are the energy flux and Unruh
temperature seen by an accelerated observer just inside the horizon.
For the so-called $f(R)$ gravity, however, Eling, Guedens, and
Jacobson~\cite{t.jac} recently have shown that  in order to derive
the corresponding field equations by using the procedure in
\cite{a8}, a treatment with nonequilibrium thermodynamics of
spacetime is needed, in which an additional entropy production term
$d\bar S$ has to be added to the Clausius relation, $\delta Q/T=
dS+d\bar S$. More recently, we have shown that a similar entropy
production term is also required for the scalar-tensor theory, while
it is not for the Lovelock gravity~\cite{CC}.

 If one identifies the black hole mass
$M$ as an energy $E$, compared to the standard first law of
thermodynamics ($dE=TdS-PdV)$, obviously a work term related to the
volume change is absent in the first law of black hole
thermodynamics $dE = TdS$.  More recently, by generalizing earlier
works~\cite{Pad}, Paranjape, Sarkar and Padmanabhan \cite{pad1} have
considered a special kind of spherically symmetric black hole
spacetimes and found that it is indeed possible to interpret
Einstein's equations as the thermodynamic identity $TdS = dE + PdV$
by considering black hole horizon as a boundary of the system, here
$P$ is the radial pressure of matter at the horizon. They have also
shown that the field equations for Gauss-Bonnet gravity and in more
general Lancos-Lovelock action in spherically symmetric space-time
can also be expressed as $TdS = dE + PdV$. These results provide a
deeper relation between the thermodynamics of horizon and Einstein's
field equations of static, spherical symmetric black hole
spacetimes.

The thermodynamical properties of the black holes horizon can be
extended to the space-time horizons other than black hole horizon.
For example, the de Sitter space time with radius $\ell$, there
exists a cosmological event horizon. This horizon, like a black hole
horizon, can be regarded as a thermodynamical system \cite{a5}
associated with the Hawking temperature $T = 1 / 2\pi\ell$ and
entropy $S = A / 4G$, where $A = 4\pi\ell^{2}$ is the cosmological
horizon area. For an asymptotic de Sitter space, like
Schwarzschild-de Sitter space-time, there still exists the
cosmological horizon which behaves like a black hole horizon with
entropy proportional to the area of the cosmological horizon and
whose Hawking temperature is given by $T = \kappa / 2\pi$, where
$\kappa$ is the surface gravity of the cosmological horizon. It is
easy to verify that the cosmological horizons of these space-times
satisfy the first law of black hole thermodynamics of the form $TdS
= -dM$ \cite{a6}, where the minus appears due to the fact that when
the black hole mass $M$ increases, the cosmological horizon radius
decreases. In this frame work, the first law $TdS = -dE$ of
cosmological horizon thermodynamics does not involve the work term
as well. By applying the first law of thermodynamics ($TdS=-dE$) to
the apparent horizon of the an $(n+1)$-dimensional
Friedman-Robertson-Walker (FRW) universe and by working out the heat
flow through the apparent horizon, one of the present authors and
Kim \cite{a11} are able to derive the Friedmann equations of the FRW
universe with any spatial curvature, during which one assumes that
the apparent horizon has temperature and entropy expressed by
\begin{equation}\label{2}
T = \frac{1}{2\pi \tilde{r}_{A}},~~~~~~ S = \frac{A}{4G},
\end{equation}
respectively, where $A$ is the area of the apparent horizon with
radius $\tilde r_A$. Also by using the entropy expression of a
static spherically symmetric black hole in the Gauss-Bonnet gravity
and in more general Lovelock gravity, they reproduce the
corresponding Friedmann equations in each gravity. The possible
extensions to the scalar-tensor gravity and $f(R)$ gravity theory
have been studied in reference \cite{a12}. In the cosmological
setting, related discussions see also, \cite{a7,a9,a10,Gau}.

On the other hand, we know that the Friedmann equations of FRW
universe are the field equations with a source of perfect fluid. In
this cosmological setup, different from the case of black hole
spacetimes discussed in \cite{pad1}, there is a well-defined concept
of pressure $P$ and energy density $\rho$. Therefore, it is required
to establish a new relationship between the first law of
thermodynamics with work term and the Friedmann equations of FRW
universe. In this respect, the authors of the present work found
\cite{acai1} that by employing Misner-Sharp energy relation inside a
sphere of radius $\tilde{r}_{A}$ of apparent horizon, it is possible
to rewrite the differential form of the Friedman equations as a
universal form $TdS = dE + WdV$, where $W = (\rho - P) / 2$. This is
nothing but the unified first law of trapping horizon firstly
suggested by Hayward~\cite{sah}. We extended this procedure to the
Gauss-Bonnet gravity and in more general Lovelock gravity, and
verified that in both cases, the Friedmann equations near apparent
horizon can also be interpreted as $TdS = dE + WdV$, where $E$ is
the total matter energy $(\rho V)$ inside the apparent horizon. The
 thermodynamics of apparent horizon in the brane world
scenario also obeys such a formula~\cite{bw}.  However, it is
interesting to investigate to what extent the horizon thermodynamics
can be developed in a spacetime against quantum gravitational
effects. A feasible way to proceed in this direction is to weigh up
the effects of higher curvature corrections to Einstein gravity,
since such corrections naturally occur due to quantum effects
\cite{ko2}. As a special case of considering the effect of higher
curvature corrections,  we will deal with in this paper the
so-called $f(R)$ gravity, whose action is an arbitrary function of
curvature scalar $R$. When $f(R)=R$, the Einstein's general
relativity is recovered.

This paper is organized as follows. In Sec.~2, we will discuss the
field equations of a general class of spherical symmetric black hole
spacetimes at the horizon and rewrite them into a form of the first
law of thermodynamics by using Hawking temperature and entropy
associated with the black hole horizon. In Sec.~3, we will discuss,
in the cosmological setup,  the corresponding field equations for
the $f(R)$ gravity. We will show that, at the apparent horizon of
FRW universe,  Friedmann equations can also be written into a form
of first law of thermodynamics. But in both cases, an additional
entropy production term will appear. Finally in Sec.~4 we will
discuss our results.


\section{Thermodynamic behavior of field equations in the black hole spacetime}

We consider a class of static, spherically symmetric spacetimes
expressed by the line element
\begin{equation}
\label{3} ds^{2} = -U(r)dt^{2} + \frac{1}{U(r)}dr^{2} +
L^{2}(r)(d\theta^{2} + \sin^{2}\theta d\phi^{2}),
\end{equation}
where $U(r)$ and $L(r)$ are two functions of $r$. We assume that the
function $U(r)$ has a simple zero at $r = a$ and $U^{\prime}(a)\neq
0$ but has a finite value at $r = a$. This defines a space-time
horizon at $r = a$. Note that $L(r)$ is a finite and continuous
function of $r$, therefore the horizon has a finite  area $A = 4\pi
L^{2}(a)$ and hence the associated entropy of the horizon is
determined by the location of the zero at $r = a$ of the function
$U(r)$.  The associated Hawking temperature with the horizon can be
obtained from the periodicity consideration of the Euclidean time.
One can obtain a Euclidean metric by Wick rotating the time
coordinate $t \rightarrow -i\tau$. This metric will be regular at
the horizon, $r = a$, if $\tau$ is taken to be an angular variable
with a period $\beta = 4\pi / U^{\prime}(a)$. This period is just
the inverse Hawking temperature $T=1/\beta$ of the black hole.  So
the non-zero surface gravity $\kappa = U^{\prime}(a) / 2$ is related
to the Hawking temperature via $T = \kappa / 2\pi$.

The components of the Einstein tensor for the metric (\ref{3}) are
given by
\begin{equation}\label{4}
G^{0}_{0} = \frac{1}{L^{2}}(-1 + U L^{\prime~2} + L( U^{\prime}
L^{\prime} + 2 U L^{\prime\prime})),
\end{equation}
\begin{equation}\label{5}
G^{1}_{1} = \frac{1}{L^{2}}(-1 + L U^{\prime} L^{\prime} + U
L^{\prime^{2}}),
\end{equation}
and
\begin{equation}\label{6}
G^{2}_{2} = G^{3}_{3} = \frac{1}{2L}(2 U^{\prime} L^{\prime} +
LU^{\prime\prime} + 2 U L^{\prime\prime}),
\end{equation}
where prime stands for the derivative with respect to $r$. It can
be seen readily that the components $G^{0}_{0}$ and $G^{1}_{1}$ of
Einstein tensor are not equal but at the spacetime horizon where
$U(a) = 0$, they become equal and are given by
\begin{equation}\label{7}
G^{0}_{0}|_{r=a} = G^{1}_{1}|_{r=a} = \frac{1}{L^{2}}(-1 + L
U^{\prime} L^{\prime}),
\end{equation}
where $U'$ and $L'$ are evaluated at $r =a$. The explicit
evaluation of curvature scalar reads
\begin{equation}\label{8}
R = -\frac{1}{L^{2}}(-2 + 2 U L^{\prime^{2}} +
L^{2}U^{\prime\prime} + 4L(U^{\prime}L^{\prime} + U
L^{\prime\prime})).
\end{equation}
At the horizon, the curvature scalar reduces to
\begin{equation}\label{9}
R|_{r = a} = -\frac{1}{L^{2}}(-2  + L^{2}U^{\prime\prime} + 4L
U^{\prime}L^{\prime}).
\end{equation}

Now let us consider the Lagrangian of the $f(R)$ gravity
\begin{equation}\label{10}
L = \frac{1}{16\pi G} f(R) + L_{m},
\end{equation}
where $f(R)$ is a continuous function of curvature scalar $R$ and
$L_{m}$ is the Lagrangian density of matter fields. Varying the
action yields corresponding equations of motion
\begin{equation}\label{11}
f_{,R}(R) R_{\mu\nu} - \frac{1}{2} g_{\mu\nu} f(R) - \nabla_{\mu}
\nabla_{\nu} f_{,R}(R) + g_{\mu\nu} \nabla^{2} f_{,R}(R) = 8\pi G
T^{m}_{\mu\nu},
\end{equation}
where $T^{m}_{\mu\nu} = \frac{-2}{\sqrt{-g}} \frac{\delta
I_{m}}{\delta g^{\mu\nu}}$ is the energy-momentum tensor for
matter fields, $\nabla$ represents the covariant derivative
defined with the Livi-Civita connection of the metric,
$R_{\mu\nu}$ is the Ricci tensor and the subscript ``$,R$''
denotes the derivative with respect to the curvature scalar $R$.
The above equations can be cast to a form
\begin{equation}\label{12}
R_{\mu\nu} - \frac{1}{2} R g_{\mu\nu} = 8\pi
G(\frac{1}{f_{,R}(R)}T^{m}_{\mu\nu} + \frac{1}{8\pi
G}T^{cur}_{\mu\nu}),
\end{equation}
where $T^{cur}_{\mu\nu}$ is a stress-energy tensor for the
effective curvature fluid and is given by
\begin{equation}\label{13}
T^{cur}_{\mu\nu} = \frac{1}{f_{,R}(R)} (\frac{1}{2}g_{\mu\nu}(f(R)
- R f_{,R}(R)) + \nabla_{\mu}\nabla_{\nu} f_{,R}(R) -
g_{\mu\nu}\nabla^{2}f_{,R}(R)).
\end{equation}
Setting $f_{,R}(R) = F(R)$ and using the relations $\nabla^{2} F =
\frac{1}{\sqrt{-g}}
 \partial_{\mu} (\sqrt{-g}g^{\mu\nu} \partial_{\nu} F)$ and
$\nabla_{\mu}\nabla_{\nu} F = \partial_{\mu}\partial_{\nu}F -
\Gamma_{\mu\nu}^{\lambda}\partial_{\lambda}F$ along with metric
(3), one can find the components of $T^{cur}_{\mu\nu}$ given by
\begin{equation}\label{14}
T^{0(cur)}_{0} = \frac{1}{F(R)}\left (\frac{1}{2}(f(R) - R F(R))
+\frac{U^{\prime}F^{\prime}}{2} -
\frac{1}{L^{2}}(2LL^{\prime}UF^{\prime} +
L^{2}U^{\prime}F^{\prime} + L^{2}UF^{\prime\prime}) \right ),
\end{equation}
and
\begin{equation}\label{15}
T^{1(cur)}_{1} = \frac{1}{F(R)}(\frac{1}{2}(f(R) - R F(R))
+(UF^{\prime\prime} + \frac{U^{\prime}F^{\prime}}{2}) -
\frac{1}{L^{2}}(2LL^{\prime}UF^{\prime} +
L^{2}U^{\prime}F^{\prime} + L^{2}UF^{\prime\prime})).
\end{equation}
Note that here the prime stands for the derivative with respect to
$r$. From equations (\ref{14}) and (\ref{15}), it is easy to see
that the components $T^{0(cur)}_{0}$ and $T^{1(cur)}_{1}$  are not
equal in general,  but at the spacetime horizon, where $U(a)$
vanishes, we have
\begin{equation}\label{16}
T^{0(cur)}_{0} = T^{1(cur)}_{1} = \frac{1}{F(R)}(\frac{1}{2}(f(R) -
R F(R)) - \frac{U^{\prime}F^{\prime}}{2}).
\end{equation}
It is evident from equations (\ref{7}) and (\ref{16}) that the
field equations are consistent with $f(R)$ gravity at the horizon
of the spherically symmetric metric (\ref{3}) provided that the
stress-energy tensor of matter fields has the form
\begin{equation}\label{17}
T^{0(m)}_{0} = T^{1(m)}_{1},
\end{equation}
at the horizon.

Note that $\nabla^{2}F(R) = \frac{1}{L^{2}}(2LL^{\prime}U
F^{\prime} + L^{2}U^{\prime}F^{\prime} + L^{2}UF^{\prime\prime})$.
At the horizon, namely,  $U(a) = 0$, one then has
$\nabla^{2}F(R)|_{r = a} = U^{\prime}(a) F^{\prime}$ and the $0-0$
component of equations of motion takes the form
\begin{equation}\label{18}
\frac{-1 + LL^{\prime}U^{\prime}}{L^{2}} = 8\pi G \left
(\frac{1}{F(R)} T^{0(m)}_{0} + \frac{1}{8\pi G
F(R)}(\frac{1}{2}(f(R) - R F(R)) - \frac{1}{2}U^{\prime}
F^{\prime}) \right ),
\end{equation}
 which can be rewritten as
\begin{equation}\label{19}
-\frac{F(R)}{2G} + \frac{U' L L'F}{2G} = 4\pi( T^{0(m)}_{0} +
\frac{1}{8\pi G}(\frac{1}{2}(f(R) - R F(R)) -
\frac{1}{2}U^{\prime} F^{\prime}))L^{2}.
\end{equation}
Note that here the curvature scalar $R$ is taken its value at the
horizon $R=R(r)|_{r=a}$.  If one has a close look at the above
equation, it throws light on the notion of horizon entropy,
temperature, and energy. So a thermodynamic interpretation of this
equation is possible. Multiplying by an infinitesimal displacement
of horizon $da$ on both sides of this equation, it is trivial to
rewrite this equation in the form
\begin{equation}\label{20}
-\frac{1}{2G}F(R) da + \frac{1}{2G} U^{\prime} F(R) LdL = 4\pi(
T^{0(m)}_{0} + \frac{1}{16\pi G} (f(R) - R F(R) - U^{\prime}
F^{\prime}) L^{2}da.
\end{equation}
Note that $F(R) LdL = \frac{1}{2}d(F(R) L^{2}) - \frac{1}{2}
L^{2}F' da$. Substituting it into the above equation, we finally
get
\begin{equation}\label{21}
-\frac{1}{2 G} F(R) da + \frac{U^{\prime}}{4 \pi} d(\frac{4 \pi
L^{2} F(R)}{4G}) = T^{0 (m)}_{0} (4 \pi L^{2} da) +
\frac{1}{4G}(f(R) - R F(R))L^{2}da.
\end{equation}
Now we first concentrate on the left hand side of the above
equation. We consider the first term $\frac{1}{2G}F(R)da =
\frac{1}{2G}f_{,R}(R)da$ of this equation and convert it to the
 limit of general relativity by setting $f(R) = R$, we get $\frac{1}{2G}da$.
  This term $\frac{1}{2G}da$ represents the energy change $dE =
\frac{1}{2G}da$ \cite{Pad,pad1} during the infinitesimal
displacement $da$ within the Einstein gravity ($E=a/2G$ is just the
Misner-Sharp energy inside the horizon~\cite{acai1}; For a
Schwarzschild black hole solution, the such defined energy $E=a/2G$
 gives the Schwarzschild mass). Naturally, we may regard  the
term $\frac{1}{2G}F(R)da  \equiv dE$ as the energy change within the
$f(R)$ gravity during a infinitesimal horizon displacement $da$. The
second term on the left hand side can be written  as $\frac{
U^{\prime}}{4\pi}d(\frac{4\pi L^{2}F(R)}{4G})$. Here one can
identify $ \frac{ U^{\prime}}{4\pi} = T$ as  the temperature of the
black hole and the term $d(\frac{4\pi L^{2}F(R)}{4G})$ is nothing
but the entropy change $dS$ of the black hole in the $f(R)$ gravity
\cite{a16, a17}. So the above equation can be rewritten as
\begin{equation}\label{22}
TdS - dE = T^{0(m)}_{0} 4\pi L^{2} da  + \frac{1}{4G}(f(R) - R
F)L^{2}da
\end{equation}
The first term of the right hand side of above equation is
$T^{0(m)}_{0} 4\pi L^{2} da$, where one can recognize $4\pi L^{2}$
as a surface area of horizon with a multiple displacement $da$ of
the horizon radius $a$,  which corresponds to a change of volume. We
write it as $4\pi L^{2} da = dV$. Since at the horizon one has
$T^{0(m)}_{0} =  T^{1(m)}_{1} \equiv P$, which represents the radial
pressure of matter fields at the horizon, therefore this term turns
out to be the work term $PdV$ against the pressure. Hence the above
equation can be written as
\begin{equation}\label{23}
TdS - dE = PdV + T \frac{A}{4G}(\frac{f(R) - R F}{U'})da,
\end{equation}
where $A= 4\pi L^2(a)$ is the area of the black hole horizon. It has
been shown in references \cite{Pad,pad1} that the field equations in
Einstein, Gauss-Bonnet and Lovelock gravities can be rewritten as
the thermodynamical identity $TdS = dE + PdV$ at the black hole
horizon of a special class of spherical symmetric black hole
spacetimes, but here it is evident from the above equation that the
field equation of $f(R)$ gravity can no longer be cast to the
thermodynamical identity $TdS = dE + PdV$ at the black hoke horizon
if one defines the horizon energy as $E=1/2G \int^a_0 F(R)da$. It is
reminiscent of the argument recently given by Eling {\it et
al.}~\cite{t.jac}. In that reference, they argued that in order to
derive the filed equations for the $f(R)$ gravity by using relation
$\delta Q= TdS$ and assuming the relation holds for all local
Rindler causal horizons through each spacetime point, here $\delta
Q$ and $T$ are the energy flux and Unruh temperature seen by an
accelerated observer just inside the horizon, and $S=f_{,R}A/4G$ is
the horizon entropy, one has to add an additional entropy term to
the Clausius relation so that $\delta Q/T=dS +d\bar S$, here the
expression of $d\bar S$ is given in \cite{t.jac}. The appearance of
the additional entropy term is interpreted as entropy production
term in the nonequilibrium thermodynamics. That is, in their
setting, the horizon thermodynamics for the $f(R)$ gravity is a
nonequilibrium one. In the spirit of \cite{t.jac}, it is then
natural to regard the last term in (\ref{23}) as the entropy
production term in our setup. In other words,  if one includes the
effects of higher curvature correction to the Einstein gravity and
the curvature scalar $R$ in general relativity is replaced by a
generic function $f(R)$ then the usual thermal behavior of the field
equations at black hole horizon \cite{Pad,pad1} will not be
sustained by the system but will shift from an equilibrium setup to
a non-equilibrium one~\cite{t.jac}. Instead of satisfying the
thermodynamical identity $TdS  = dE + PdV$  at the black hole
horizon, we have to put in the effects of higher curvature
correction to the horizon entropy grown up internally because of the
system being out of equilibrium~\cite{t.jac}. So one may have a
balance thermal identity for higher curvature gravity theory
\begin{equation}\label{24}
dE = TdS - PdV + Td\bar{S}
\end{equation}
where $Td\bar{S}$ is just the last term of (\ref{23}), which is the
entropy production term added to balance the inequality $TdS > dE +
PdV$. We note that the entropy production term for the  $f(R)$
gravity has the form
\begin{equation}\label{25}
d\bar{S} = \frac{A}{4G}(\frac{RF-f(R) }{U'})da,
\end{equation}
in our setup of spherically symmetric black hole horizon.  It is
evident from this equation that the entropy production term vanishes
at the limit $f(R) = R$ of the Einstein's gravity,  and the usual
thermal behavior $TdS = dE + PdV$ of the field equation at the
horizon is recovered.

Clearly the above interpretation of nonequilibrium thermodynamics of
black hole horizon is not unique. One may combine the last term in
(\ref{21}) to the first term so that one can define a new energy
associated with the black hole horizon as
 \begin{equation}
 \label{26}
 d\bar E= \frac{1}{2G}Fda +\frac{1}{4G}(f(R)-RF)L^2 da.
 \end{equation}
 In this way, we can rewrite (\ref{21}) to the standard first law of
 thermodynamics
 \begin{equation}
 \label{in1}
  d\bar E=TdS-PdV.
  \end{equation}
Take an example, let us consider a constant curvature black hole
solution of the $f(R)$ gravity without matter fields. In that case,
we have from (\ref{11}) that $f_{,R}R_0=2f(R_0)$, where the $R_0$ is
the curvature scalar of the black hole solution. The field equations
(\ref{11}) then can be cast to the standard Einstein's equations
with an effective cosmological constant $\Lambda\equiv
3/\ell^2=R_0/4$. Thus we have a simple Schwarzschild-de Sitter
(anti-de Sitter) black hole solution depending on the sign of the
effective cosmological constant, which has the metric functions,
$L(r)=r$ and
\begin{equation}
U(r)=1-\frac{2GM}{r}-\frac{r^2}{\ell^2},
\end{equation}
where $M$ is the mass of the black hole in the Einstein gravity.  In
terms of the black hole horizon radius $U(r)|_{r=a}=0$, the mass can
be expressed by
\begin{equation}
\label{in3}
 M=\frac{a}{2G}\left (1-\frac{a^2}{\ell^2}\right).
 \end{equation}
 On the other hand, it is interesting to note that integrating
 (\ref{26}) yields
 \begin{equation}
 \bar E= F(R_0)M.
 \end{equation}
 Clearly, the black hole mass $M$ does not satisfy the equation
 (\ref{in1}) with $P=0$, but $\bar E$ does.  Therefore defining
 $\bar E$ as the horizon energy of the black hole is reasonable.

\section{Thermodynamics of apparent horizon of FRW universe}

In the previous section we have discussed the behavior of the field
equations for a static, spherically symmetric black hole spacetimes
at the black hole horizon. In this section we generalize the
thermodynamics of black hole horizon to a dynamical apparent horizon
in a FRW universe within the $f(R)$ gravity. We consider an
$(n+1)$-dimensional, spatially homogenous and isotropic  FRW
universe described by the  metric
\begin{equation}\label{27}
 ds^{2} = -dt^2 + a(t)^{2}(\frac{dr^{2}}{1-k r^{2}} +
r^{2}d\Omega^{2}_{n-1}),
\end{equation}
where $a(t)$ is the scale factor of the universe with $t$ being
cosmic time and $d\Omega^{2}_{n-1}$ is the metric of
$(n-1)$-dimensional sphere with unit radius and the spatial
curvature constant $k = 1$, $0$ and $-1$ correspond to a closed,
flat and open universe, respectively. Using the spherical symmetry,
the metric (\ref{27}) can be rewritten as
\begin{equation}\label{28}
ds^{2} = h_{ab}dx^{a}dx^{b} + \tilde{r}^{2}d\Omega^{2}_{n-1},
\end{equation}
where $\tilde{r}= a(t)r$ and $x^{0}=t$, $x^{1}=r$ and the two
dimensional metric $h_{ab} = {\rm diag}(-1, a^{2}/1-kr^{2})$. The
dynamical apparent horizon is determined by the relation
$h^{ab}\partial_{a}\tilde{r}\partial_{b}\tilde{r}=0$, which implies
that the vector $\nabla \tilde{r}$ is null on the apparent horizon
surface. The explicit evaluation of the apparent horizon gives the
apparent horizon radius
\begin{equation}\label{29}
\frac{1}{\tilde{r}_{A}^{2}} = H^{2}+\frac{k}{a^{2}},
\end{equation}
where $H$ denotes for the Hubble parameter.  The associated
temperature $T = \kappa / 2\pi$ with the apparent horizon is defined
through the surface gravity
\begin{equation}\label{30}
\kappa = \frac{1}{2\sqrt{-h}}
\partial_{a}(\sqrt{-h}h^{ab}\partial_{b}\tilde{r}),
\end{equation}
which gives
\begin{equation}\label{31}
\kappa = -\frac{1}{\tilde{r}_{A}}(1 -
\frac{\dot{\tilde{r}}_{A}}{2H\tilde{r}_{A}}).
\end{equation}
From this equation one can see that when $\dot{\tilde{r}}_{A} <
2H\tilde{r}_{A}$, the surface gravity $\kappa$ is negative. If one
still uses $T = \kappa / 2\pi$ to define a temperature of the
apparent horizon, we obtain a negative temperature! This case is
quite similar to the case of cosmological horizon for the
Schwarzschild-de Sitter spacetime. In this case, one should use $T =
|\kappa| / 2\pi$ to define the temperature of the apparent horizon.

The total matter energy inside a sphere of radius $\tilde{r}_{A}$
 of the apparent horizon is given by
\begin{equation}\label{32}
E = \Omega_{n}\tilde{r}_{A}^{n}\rho,
\end{equation}
where $V = \Omega_{n}\tilde{r}_{A}^{n}$ is the volume of the sphere
within the apparent radius, and $\rho$ is the energy density of the
perfect fluid in the universe. We consider FRW universe as a
thermodynamical system with apparent horizon surface as a boundary
of the system. In general the radius of the apparent horizon
$\tilde{r}_{A}$ is not constant but changes with time. Let
$d\tilde{r}_{A}$ be an infinitesimal change in radius of the
apparent horizon during a time of interval $dt$. This small
displacement $d\tilde{r}_{A}$ in the radius of apparent horizon will
cause a small change $dV$ in the volume $V$ of the apparent horizon.
Since the energy $E$ inside the apparent horizon of FRW universe is
directly related with the apparent radius $\tilde{r}_{A}$ therefore
an infinitesimal change $d\tilde{r}_{A}$ will result in a change
$dE$ in energy $E$ and is given by
\begin{equation}\label{33}
dE = n\Omega_{n}\tilde{r}_{A}^{n}\rho d\tilde{r}_{A} - n\Omega_{n}
\tilde{r}_{A}^{n}(\rho + P) H dt.
\end{equation}
Here, we have utilized the continuity equation (\ref{35}).

Now we consider the field equations (\ref{12}), the stress-energy
tensor $T^{m}_{\mu\nu}$ is taken to be the one of the perfect fluid
with  time dependent energy density $\rho(t)$ and pressure $P(t)$
\begin{equation}\label{34}
T^{m}_{\mu\nu} = (\rho + P)U_{\mu}U_{\nu} + P g_{\mu\nu},
\end{equation}
where $U_{\mu}$ is the four velocity of the fluid. The conservation
of the stress-energy tensor,$T^{\mu\nu(m)}_{;\nu} = 0$, leads to the
continuity equation
\begin{equation}\label{35}
\dot{\rho} + n H(\rho + P) = 0,
\end{equation}
where the dot represents derivative with respect to cosmic time $t$.

Using the relations $\nabla^{2} F = \frac{1}{\sqrt{-g}}
 \partial_{\mu} (\sqrt{-g}g^{\mu\nu} \partial_{\nu} F)$ and
$\nabla_{\mu}\nabla_{\nu} F = \partial_{\mu}\partial_{\nu}F -
\Gamma_{\mu\nu}^{\lambda}\partial_{\lambda}F$ along with the
$(n+1)$-dimensional metric (\ref{27}) of  the FRW universe, one can
find the components of $T^{cur}_{\mu\nu}$
\begin{eqnarray}\label{37}
T^{cur}_{00} &=& -\frac{1}{F(R)}(\frac{1}{2}(f(R) - R F(R)) + nH
F_{,R}(R) \dot{R}), \\
T^{cur}_{ii} & =& \frac{1}{F(R)}\left (\frac{1}{2}(f(R) - R F(R)) +
F_{,R}(R) \ddot{R} + F_{,R,R}(R) \dot{R}^{2} + (n-1)H F_{,R}(R)
\dot{R} \right ) g_{ii},
\end{eqnarray}
where $F(R)=df/dR$, $F_{,R}=dF/dR$ and $F_{,R,R}=d^2F/dR^2$.  The
field equations become
\begin{eqnarray}
 \label{39}
&& H^{2} + \frac{k}{a^{2}} = \frac{16\pi G}{n(n-1)} \left (
\frac{1}{F(R)} \rho - \frac{1}{8\pi G F(R)}(\frac{1}{2}(f(R) - R
F(R)) + nH
F_{,R}(R) \dot{R} \right ), \\
&& \dot{H} - \frac{k}{a^{2}} = - \frac{8\pi G}{(n-1)F(R)}\left
((\rho + P) + \frac{1}{8\pi G}\left (d(F_{,R}(R) \dot{R}) - H
f^{\prime\prime}(R) \dot{R} \right ) \right ), \label{40}
\end{eqnarray}
where $d(F_{,R}(R)) \equiv d(F_{,R}(R))/dt$.  These are two
Friedmann equations for the $f(R)$ gravity \cite{b1}. Note that, as
mentioned in the previous section, in this gravity theory the black
hole entropy has a relation to its horizon area $S = \frac{A}{4G}
F(R)|_{\rm Horizon}$. We assume it also holds for the apparent
horizon.  Using relation $H^{2} + \frac{k}{a^{2}} =
\frac{1}{\tilde{r}_{A}^{2}}$ and taking differential of it, one gets
\begin{equation}\label{41}
-\frac{1}{\tilde{r}_{A}^{3}} d\tilde{r}_{A} = H(\dot{H} -
\frac{k}{a^{2}})dt.
\end{equation}
Substituting it into (\ref{40}) and then multiplying in both sides
with a factor $n\Omega_{n}\tilde{r}_{A}^{n-3}$, we reach
\begin{equation}\label{42}
\frac{1}{2\pi \tilde{r}_{A}}(n(n-1)\Omega_{n}\tilde{r}_{A}^{n-2}
F(R) d\tilde{r}_{A}) / 4G = n\Omega_{n}\tilde{r}_{A}^{n}H \left
((\rho + P) + \frac{1}{8\pi G}(d(F_{,R}(R) \dot{R}) - H
F_{,R}\dot{R}) \right )dt.
\end{equation}
Now consider the differential
\begin{equation}\label{43}
d(n\Omega_{n}\tilde{r}_{A}^{n-1}F(R) / 4G) =
n(n-1)\Omega_{n}\tilde{r}_{A}^{n-2}F(R) d\tilde{r}_{A} / 4G +
n\Omega_{n}\tilde{r}_{A}^{n-1}F_{,R}(R)\dot{R}dt / 4G,
\end{equation}
and substitute it into  (\ref{42}), one gets
\begin{eqnarray}\label{44}
\frac{1}{2\pi \tilde{r}_{A}}d(\frac{n\Omega_{n}\tilde{r}_{A}^{n-1}
F(R)}{4G}) &=& n\Omega_{n}\tilde{r}_{A}^{n}H \left ((\rho + P) +
\frac{1}{8\pi G}(d(F_{,R}(R) \dot{R}) - H F_{,R}\dot{R}) \right
)dt  \nonumber \\
 && +
\frac{n\Omega_{n}\tilde{r}_{A}^{n-2}F_{,R}\dot{R}dt}{8\pi G}.
\end{eqnarray}
Multiplying  both sides of this equation by a factor $-(1 -
\dot{\tilde{r}}_{A} / 2H \tilde{r}_{A})$ and identifying that the
surface gravity $\kappa = (-1 / \tilde{r}_{A})(1 -
\dot{\tilde{r}}_{A} / 2H \tilde{r}_{A})$ at the apparent horizon,
one can obtain
\begin{eqnarray}\label{45}
\frac{\kappa}{2\pi} d(\frac{A F(R)}{4 G}) &=& -(1 -
\frac{\dot{\tilde{r}}_{A}}{2H
\tilde{r}_{A}})n\Omega_{n}\tilde{r}_{A}^{n}H \left ((\rho + P) +
\frac{1}{8\pi G}(d(F_{,R}(R) \dot{R}) - H
F_{,R}\dot{R} \right )dt \nonumber \\
&& -(1 - \frac{\dot{\tilde{r}}_{A}}{2H \tilde{r}_{A}})
\frac{n\Omega_{n}\tilde{r}_{A}^{n-2}F_{,R}\dot{R}}{8\pi G}dt,
\end{eqnarray}
where $A = n\Omega_{n}\tilde{r}_{A}^{n-1}$ is the area of the
apparent horizon and $\Omega_{n} = \pi^{n/2} / \Gamma(n/2 + 1)$
being the volume of an $n$-dimensional unit ball.  In addition,
let us note that all quantities in the above equation are
evaluated at the apparent horizon. Recognizing $\kappa / 2\pi$ as
the temperature $T$ of the apparent horizon and the quantity
inside the parentheses on the left hand side of this equation, is
just the entropy $S = \frac{A}{4G} F(R)|_{\tilde{r}_{A}}$  of the
apparent horizon in the $f(R)$ gravity.  Hence, having identified
temperature $T$ and entropy $S$ , the above equation can be
written as
\begin{eqnarray}\label{46}
T dS  &=& -n\Omega_{n}\tilde{r}_{A}^{n}(\rho + P)H dt +
\frac{n}{2}\Omega_{n}\tilde{r}_{A}^{n - 1}(\rho + P)d\tilde{r}_{A}
 \nonumber \\
 && + T \frac{A}{4 G}\left (H \tilde{r}_{A}^{2} (d(F_{,R}(R)
\dot{R}) - H F_{,R}(R) \dot{R}) + F_{,R}(R)\dot{R} \right )dt.
\end{eqnarray}
Now we turn to the total matter energy $E =
\Omega_{n}\tilde{r}_{A}^{n}\rho$  inside the apparent horizon. Using
(\ref{33}), the above equation becomes
\begin{equation}\label{47}
TdS -dE = -\frac{1}{2}(\rho - P)d(\Omega_{n}\tilde{r}_{A}^{n})+T
\frac{A}{4 G}\left (H \tilde{r}_{A}^{2} (d(F_{,R}(R) \dot{R}) - H
F_{,R}(R) \dot{R}) + F_{,R}(R)\dot{R} \right )dt.
\end{equation}
Note that $ (\rho -P) / 2 = W$ is the work density~\cite{sah}. Hence
the equation (\ref{47}) can be further rewritten as
\begin{equation}\label{48}
dE = TdS + W dV -T \frac{A}{4 G}\left (H \tilde{r}_{A}^{2}
(d(F_{,R}(R) \dot{R}) - H F_{,R}(R) \dot{R}) + F_{,R}(R)\dot{R}
\right )dt.
\end{equation}
In the cosmological setting, we have already shown~\cite{acai1}
that the field equations obey the universal form $dE = TdS + WdV$
at the apparent horizon of FRW universe in Einstein, Gauss-Bonnet
and Lovelock gravities. But it no longer holds for the
scalar-tensor theory~\cite{CC}. Here we see from the equation
(\ref{48}) that the universal form $dE = TdS + WdV$ also does not
hold for the $f(R)$ gravity at the apparent horizon of FRW
universe because of the appearance of the additional term $T
\frac{A}{4 G}\left (H \tilde{r}_{A}^{2} (d(F_{,R}(R) \dot{R}) - H
F_{,R}(R) \dot{R}) + F_{,R}(R)\dot{R} \right )dt$.  Of course when
$f(R)=R$, the additional term is absent, the usual result for the
Einstein gravity is recovered~\cite{acai1}. In the spirit of the
argument of \cite{t.jac}, once again, the additional term can be
interpreted as follows. This additional term is developed
internally in the system as a result of being out of equilibrium.
 In order to get a
balance relation for a universal form, one may take
\begin{equation}\label{49}
dE = TdS + WdV + Td\bar{S}
\end{equation}
The term $d\bar{S}$ is the entropy production term grown up
internally due to the non-equilibrium setup within the $f(R)$
gravity at the apparent horizon of FRW universe. Comparing
equation (\ref{48}) and (\ref{49}), we get
\begin{equation}\label{50}
d\bar{S} =-\frac{A}{4 G}\left (H \tilde{r}_{A}^{2} (d(F_{,R}(R)
\dot{R}) - H F_{,R}(R) \dot{R}) + F_{,R}(R)\dot{R} \right )dt.
\end{equation}
If one further defines $d\tilde{S} = d(S + \bar{S})$ as an
effective entropy change during the infinitesimal displacement
$d\tilde{r}_{A}$ of the apparent horizon of FRW universe in
interval $dt$, the above equation (\ref{49}) can be rewritten as
\begin{equation}\label{51}
dE = Td\tilde{S }+ WdV,
\end{equation}
where $\tilde{S} = S + \bar{S}$ is the effective entropy
associated with the apparent horizon of FRW universe.

\section{Conclusion and Discussion}

Black hole thermodynamics implies that there must exist some
relation between gravity theory and the laws of thermodynamics.
Indeed, it has been shown that at the horizon of spherical symmetric
black hole spacetime, field equations can be written into the form
of the first law of thermodynamics, $dE=TdS-PdV$, not only for the
Einstein gravity, but also Gauss-Bonnet gravity and more general
Lovelock gravity~\cite{Pad,pad1}. In the cosmological setting,
applying the Clausius relation $\delta Q=TdS$ to the apparent
horizon of a FRW universe, one can derive corresponding Friedmann
equations of the universe with any spatial curvature in the
Einstein, Gauss-Bonnet and Lovelock gravity theories~\cite{a11}.
Further we have shown that at the apparent horizon, the field
equations of gravity can also be cast to a similar form of the first
law, $dE=TdS+WdV$, in the Einstein, Gauss-Bonnet, and Lovelock
gravity theories~\cite{acai1}. To what extent does such a formulism
hold? In the present paper we have discussed the so-called $f(R)$
gravity.

In the setup of static, spherically symmetric black hole spacetime,
we have shown that the field equations of the $f(R)$ gravity can be
rewritten as $dE  = TdS - PdV + Td\bar{S}$, at the black hole
horizon. Here $E=1/2G \int^a_0 F(R)da$ is regarded as the horizon
energy; when $f(R)=R$, the horizon energy reduces to the one given
in \cite{Pad} for Einstein gravity. $T$ is the Hawking temperature
of the black hole and $S=F(R)A/4G$ is the horizon entropy of the
black hole, and $P$ is the radial pressure of matter fields at the
horizon. The additional term $d\bar S$ given in (\ref{25}) can be
regarded as an entropy production term due to a nonequilibrium
thermodynamics setup~\cite{t.jac}, where it is argued that in the
setup of a class of Rindler causal horizons, in order to derive the
field equations for the $f(R)$ gravity, an entropy production term
has to be added to the Clausius relation $\delta Q =TdS +Td_i S$.
However, in our case, we have argued that the additional term can be
combined to the horizon energy term by defining a new horizon energy
(\ref{26}). In this way, the field equations of $f(R)$ gravity can
be cast into the standard form of the first law of thermodynamics
(\ref{in1}): $d\bar E=TdS-PdV$, which reduces to the first law of
black hole thermodynamics when $P=0$.

In the setup of cosmology of FRW universe, compared to the cases
of Einstein gravity, Gauss-Bonnet gravity and Lovelock gravity, we
have found that an additional term appears when the Friedmann
equations of the $f(R)$ gravity are rewritten into a form of the
first law of thermodynamics at the apparent horizon of the
universe, $dE=TdS+WdV + Td\bar S$. As usual, here $E$ is the total
energy of matter inside the apparent horizon, $T$ and $S$ are the
associated temperature and entropy with the horizon, and
$W=(\rho-P)/2$ is the work density. The additional term in this
case is given by (\ref{50}). Once again, the additional term can
be interpreted as the entropy production term associated with the
apparent horizon in the nonequilibrium thermodynamics within the
$f(R)$ gravity in the spirit of \cite{t.jac}.

The authors of \cite{t.jac} used the modified Clausius relation
\begin{equation}\label{55}
 \delta Q= TdS + Td_{i}S,
\end{equation}
to a Rindler horizon of spacetime with the assumption that the
horizon has a  temperature $T = 1/ 2\pi$ and entropy $S =  A
F(R)/4G$, where $A$ is the horizon area.  They defined the heat as
the mean flux of the boost energy current of matter across the
horizon
\begin{equation}
\delta Q = \int T_{ab}\chi^{a}d\Sigma^{b}.
\end{equation}
In this frame work, in order to get the correct field equations of
the $f(R)$ gravity, they worked out that the entropy production term
$d_{i}S$ is
\begin{equation}
\label{57} d_{i}S = -\frac{3}{8G} \int F(R) \theta^{2} \lambda
d\lambda d^{2}A
\end{equation}
where $\theta = d(\ln d^{2}A) / d\lambda$ is the expansion of the
congruence of null geodesics generating the horizon and $\lambda$
is an affine parameter.

We note from (\ref{25}), (\ref{50}) and (\ref{57}) that in the
different settings, the additional entropy term has a different from
for the $f(R)$ gravity. The same happens in the scalar-tensor
gravity~\cite{CC}. It would be of great interest to further
understand this issue. In addition, why does such an additional term
appears for the $f(R)$ gravity and scalar-tensor gravity, while it
does not in the Einstein's gravity, Gauss-Bonnet gravity and more
general Lovelock gravity. This is a quite interesting question,
which deserves further investigation. This might be related to the
observation that the field equations for the Einstein's gravity,
Gauss-Bonnet gravity and Lovelock gravity can also be derived from a
holographic surface term~\cite{Pad2}.

\section*{Acknowledgments}
We thank L.M. Cao, B. Wang and R.K. Su for useful discussions.
 This work was supported in part by a grant from Chinese
Academy of Sciences, and grants from NSFC, China (No. 10325525 and
No. 90403029).



\begin{thebibliography}{99}
{\small

\bibitem{a3} J. M. Bardeen, B. Carter and S. W. Hawking, Commun. Math. Phys. \textbf{31}, 161
(1973).

\bibitem{a1} S. W. Hawking, Commun. Math. Phys., {\bf{43}}, 199
(1975).
\bibitem{a2} J. D. Bekenstein, Phys. Rev. D\textbf{7}, 2333 (1973).

\bibitem{a8} T. Jacobson, Phys. Rev. Lett. \textbf{75}, 1260 (1995).

\bibitem{t.jac} C. Eling, R. Guedens, and T. Jacobson,
Phys.Rev.Lett.\textbf{96}, 121301 (2006) [arXiv:gr-qc/0602001].

\bibitem{CC}R.~G.~Cai and L.~M.~Cao,
  arXiv:gr-qc/0611071.



\bibitem{Pad}T. Padmanabhan, Class. Quant. Grav. {\bf 19}, 5387
(2002)[arXiv: gr-qc/0204019]; Phys. Rept. {\bf 406}, 49
(2005)[arXiv: gr-qc/0311036]; T.~Padmanabhan,
  arXiv:gr-qc/0606061.

\bibitem{pad1} A. Paranjape, S. Sarkar, and T. Padmanabhan,
Phys. Rev. D\textbf{74}, 104015 (2006) [arXiv: hep-th/0607240].



\bibitem{a5} G. W. Gibbons and S. W. Hawking, Phys. Rev. D\textbf{15}, 2738
(1977).
\bibitem{a6} R. G. Cai, Nucl. Phys. \textbf{B628}, 375 (2002) [arXiv: hep-th/0112253];
R. G. Cai, Phys. Lett.B \textbf{525}, 331
(2002)[arXiv:hep-th/0111093].


\bibitem{a11} R. G. Cai and S. P. Kim, JHEP
\textbf{0502}, 050 (2005)[arXiv:hep-th/0501055].
\bibitem{a12} M.
Akbar and R. G. Cai, Phys. Lett. \textbf{B635}, 7 (2006)
[arXiv:hep-th/0602156].

\bibitem{a7} A. V. Frolov and L. Kofman, JCAP \textbf{0305},
009(2003) [hep-th/0212327].

\bibitem{a9} U. K. Danielsson, Phys. Rev.
D\textbf{71}, 023516(2005) [arXiv: hep-th/0411172].
\bibitem{a10}
R. Bousso, Phys. Rev. D\textbf{71}, 064024 (2005) [arXiv:
hep-th/0412197].

\bibitem{Gau}G.~Calcagni,
  JHEP {\bf 0509}, 060 (2005)
  [arXiv:hep-th/0507125].

\bibitem{acai1} M. Akbar and R. G. Cai, arXiv: hep-th/0609128.

\bibitem{sah} S. A. Hayward,
Class. Quant. Grav.\textbf{15}, 3147 (1998) [Archive:
gr-qc/9710089]; S. A. Hayward, S. Mukohyana, and M. C. Ashworth,
Phys. Lett. A\textbf{256}, 347 (1999).

\bibitem{bw} R.G. Cai and L.M. Cao, arXiv: hep-th/0612144.

\bibitem{ko2} N. D. Birrel and P. C. W. Davies, {\it Quantum Fields in
Curved Space} (Cambridge University Press, Combridge, 1982).

\bibitem{a16} R. M. Wald, Phys. Rev. D\textbf{48}, 3427(1993).

\bibitem{a17} G. Cognola, E. Elizalde, S. Nojiri, S. D. Odintsov,
and S. Zerrbini, JCAP \textbf{0502}, 010 (2005); I.~Brevik,
S.~Nojiri, S.~D.~Odintsov and L.~Vanzo,
  Phys.\ Rev.\ D {\bf 70}, 043520 (2004)
  [arXiv:hep-th/0401073].
  
\bibitem{b1} S. Capozziello, V. F. Cardone, and A. Troisi, Phys.
Rev. \textbf{D71}, 043503 (2005).

\bibitem{Pad2}A.~Mukhopadhyay and T.~Padmanabhan,
  arXiv:hep-th/0608120.

}

\end{thebibliography}
\end{document}